\documentclass[11pt,a4paper]{article}
\usepackage[utf8]{inputenc}
\usepackage{amsmath,amssymb,amsfonts}
\usepackage{graphicx}
\usepackage{booktabs}
\usepackage{hyperref}
\usepackage{cite}
\usepackage[margin=1in]{geometry}

\title{\textbf{Dynamic Democratic Consent: A System Dynamics Framework for Public Trust, Political Communication, and Policy Feedback}}
\author{\textbf{Muhammad Sukri Bin Ramli}\\
Asia School of Business, Kuala Lumpur, Malaysia\\
\texttt{m.binramli@sloan.mit.edu}}
\date{August 28, 2026}

\begin{document}

\maketitle

\begin{abstract}
Political communication is often analyzed through event-based models that give limited attention to feedback, delay, saturation, and changing public trust. This conceptual paper proposes the Dynamic Democratic Consent framework, integrating selected principles from Edward Bernays' account of public relations with the S-E-E-D behavior modes of Snowball growth, Equilibrium seeking, Elastic adjustment, and feedback-loop Dominance. The framework extends the conventional resource categories of manpower, mindpower, and money by treating algorithmic media infrastructure as a conditional fourth dimension. It maps an eight-stage communication-planning process onto reinforcing and balancing feedback structures, perception delays, demographic capacity, and institutional trust. Simulated reference modes illustrate how political-support momentum may encounter electoral saturation, how delayed perception and corrective response may produce oscillatory adjustment, and how divergence between communication and policy delivery may weaken institutional trust and contract effective support capacity. A four-stage diagnostic process is proposed for describing trajectories, identifying feedback structures, designing policy interventions, and examining potential behavioral shifts. The framework is exploratory rather than predictive. Empirical estimation, comparative case testing, sensitivity analysis, and ethical evaluation are required before it can support applied political analysis.
\end{abstract}

\section{Introduction and Theoretical Synthesis}
Many conventional campaign analyses rely on event-centered or comparatively linear representations of political persuasion~\cite{bennett2008, zaller1992, ramli2026}. These models treat voter persuasion as a sequence of isolated inputs, assuming that campaign advertising, debate performances, or policy announcements produce proportional and immediate shifts in electoral support~\cite{bernays1955, mccombs1972}. Such static assumptions create a persuasion-feedback gap~\cite{ramli2026}. In complex democratic ecosystems, public opinion does not respond linearly; rather, it functions as an interconnected system governed by self-reinforcing acceleration, structural balancing constraints, material information delays, and dynamic saturation boundaries~\cite{entman1993, sterman2000, ramli2026}. Failing to account for these systemic properties leaves political organizations vulnerable to voter fatigue, political volatility, and erosion of institutional credibility~\cite{bernays1955, hetherington2005}.

To resolve these analytical limitations, this paper establishes a novel theoretical synthesis termed the \textbf{Dynamic Democratic Consent (DDC)} framework. DDC critically integrates selected elements of Edward Bernays' foundational doctrine, which asserts that the continuous, scientific cultivation of public consent is an indispensable component of democratic governance predicated on ``deeds explained by words''~\cite{bernays1955}, with the quantitative feedback modeling of the S-E-E-D framework in system dynamics~\cite{forrester1961, sterman2000, ramli2026}. Where Bernays provided the qualitative blueprint for structural persuasion and public relations counsel, the S-E-E-D framework supplies the mathematical architecture to conceptualize how positive reinforcing loops ($R_1$), negative balancing loops ($B_1, B_2$), perception adjustment time constants ($AT_p$), and carrying capacity limits ($K_{\text{dem}}$) interact over time~\cite{bernays1955, sterman2000, ramli2026}. By formalizing these dynamics, DDC frames political support acquisition and retention not as an art of speculative intuition or short-term publicity, but as a feedback-informed analytical framework for analyzing dynamic political support structures~\cite{bernays1955, pierson1993, ramli2026}.

The central research question guiding this inquiry is: \textit{How do feedback-loop structures, communication delays, policy delivery, and institutional trust interact to shape the evolution of public support in democratic political systems?} To address this question, we examine the structural mechanisms that govern voter mobilization, map campaign execution onto system dynamics behavior modes, and provide a diagnostic framework capable of identifying leverage points within complex political networks~\cite{bernays1955, sterman2000, ramli2026}.

\subsection{Literature Synthesis and Research Gap}
Political communication scholarship has extensively investigated how media messaging, framing, and agenda-setting influence public opinion and electoral choice~\cite{entman1993, mccombs1972, zaller1992}. Recent studies further highlight how computational propaganda, automated bots, and algorithmic micro-targeting accelerate message distribution and fragment audience attention~\cite{vosoughi2018, woolley2016}. Concurrently, political science research on policy feedback demonstrates that government actions and administrative outcomes actively reshape political identities, citizen trust, and subsequent electoral demands over extended time horizons~\cite{hetherington2005, pierson1993}.

In parallel, the field of system dynamics provides rigorous conceptual and mathematical methods for modeling feedback structures, material accumulations, non-linear interactions, and delay-induced oscillations across industrial and social systems~\cite{forrester1961, sterman2000}. Although system dynamics has been applied to organizational management and public policy, its formal integration with political communication planning and public relations theory remains underdeveloped~\cite{sterman2000, ramli2026}.

Despite substantial research within each field, comparatively limited attention has been given to integrating political communication planning, policy feedback, institutional trust, and system-dynamics behavior modes within a common analytical architecture~\cite{bernays1955, sterman2000, ramli2026}. The contribution of this paper lies in organizing these mechanisms into a shared system-dynamics framework and mapping them to a structured communication-planning process. The resulting framework distinguishes among momentum, saturation, delayed adjustment, trust erosion, and changes in loop dominance, thereby generating propositions that can be evaluated in subsequent empirical work.

\section{The Expanded 4M Political Communication Resource Paradigm}
In \textit{The Engineering of Consent}, Bernays identified three fundamental resource constraints that govern all strategic public relations activities: Manpower, Mindpower, and Money~\cite{bernays1955}. While this triad effectively captured the operational limitations of twentieth-century mass media environments, the emergence of digital communications requires an updated paradigm~\cite{bernays1955, woolley2016, ramli2026}. We expand this resource base into the \textbf{4M Political Communication Resource Paradigm}, introducing \textit{Media (Algorithmic Infrastructure)} as the crucial fourth dimension~\cite{bernays1955, ramli2026}.

\begin{table}[ht]
\centering
\caption{The Expanded 4M Political Communication Resource Paradigm}
\label{tab:4m_paradigm}
\begin{tabular}{p{2.2cm}p{3.2cm}p{3.8cm}p{4.2cm}}
\toprule
\textbf{Resource} & \textbf{Bernaysian Concept} & \textbf{System Dynamics (S-E-E-D)} & \textbf{Modern Political Function} \\
\midrule
\textbf{Manpower} & Field staff, precinct cadres, internal and external publics~\cite{bernays1955}. & System stock capacity and physical bandwidth~\cite{sterman2000, ramli2026}. & Ground-game mobilization, precinct coverage, turnout operations~\cite{bernays1955, ramli2026}. \\
\textbf{Mindpower} & Strategic intelligence, framing, research~\cite{bernays1955}. & Problem formulation and loop identification~\cite{sterman2000, ramli2026}. & Polling analysis, cognitive framing, policy alignment~\cite{bernays1955, entman1993, ramli2026}. \\
\textbf{Money} & Budgetary limits and capital allocation~\cite{bernays1955}. & Inflow rate governing operational scale~\cite{sterman2000, ramli2026}. & Multi-channel advertising, infrastructure funding~\cite{bernays1955, ramli2026}. \\
\textbf{Media (4th M)} & Algorithmic networks and distribution platforms~\cite{bernays1955}. & Conditional dynamic amplifiers influencing time lags ($AT_p$)~\cite{ramli2026}. & Micro-targeting, network echo chambers, message amplification~\cite{vosoughi2018, woolley2016, ramli2026}. \\
\bottomrule
\end{tabular}
\end{table}

Manpower represents the physical human deployment bandwidth of the political organization, encompassing internal publics (staff, executives, volunteer networks) and external operational cadres required to execute ground campaigns~\cite{bernays1955, ramli2026}. Mindpower constitutes the analytical and creative capacity required to formulate clear objectives, interpret raw public opinion data, align corporate or party goals with broader civic interest, and design high-leverage policy interventions~\cite{bernays1955, entman1993, ramli2026}. Money functions as the monetary inflow rate that dictates the overall scale of campaign operations and sets hard resource ceilings~\cite{bernays1955, ramli2026}. 

Media, as the fourth M, fundamentally alters the physical dynamics of political messaging~\cite{bernays1955, woolley2016, ramli2026}. Compared with communication systems dominated by slower and more centralized distribution channels, algorithmic media infrastructure may increase transmission speed, network reach, and feedback intensity~\cite{bernays1955, vosoughi2018, ramli2026}. Algorithmic platforms can compress the perception adjustment time constant ($AT_p$), enabling rapid micro-targeting, content amplification, and automated sentiment reinforcement~\cite{woolley2016, ramli2026}. However, Media functions as a conditional dynamic multiplier rather than a deterministic engine: its amplifying potential is moderated by platform architecture, audience network fragmentation, contextual noise, and opposition counter-messaging~\cite{bennett2008, vosoughi2018}.

\begin{figure}[ht]
\centering
\includegraphics[width=0.85\linewidth]{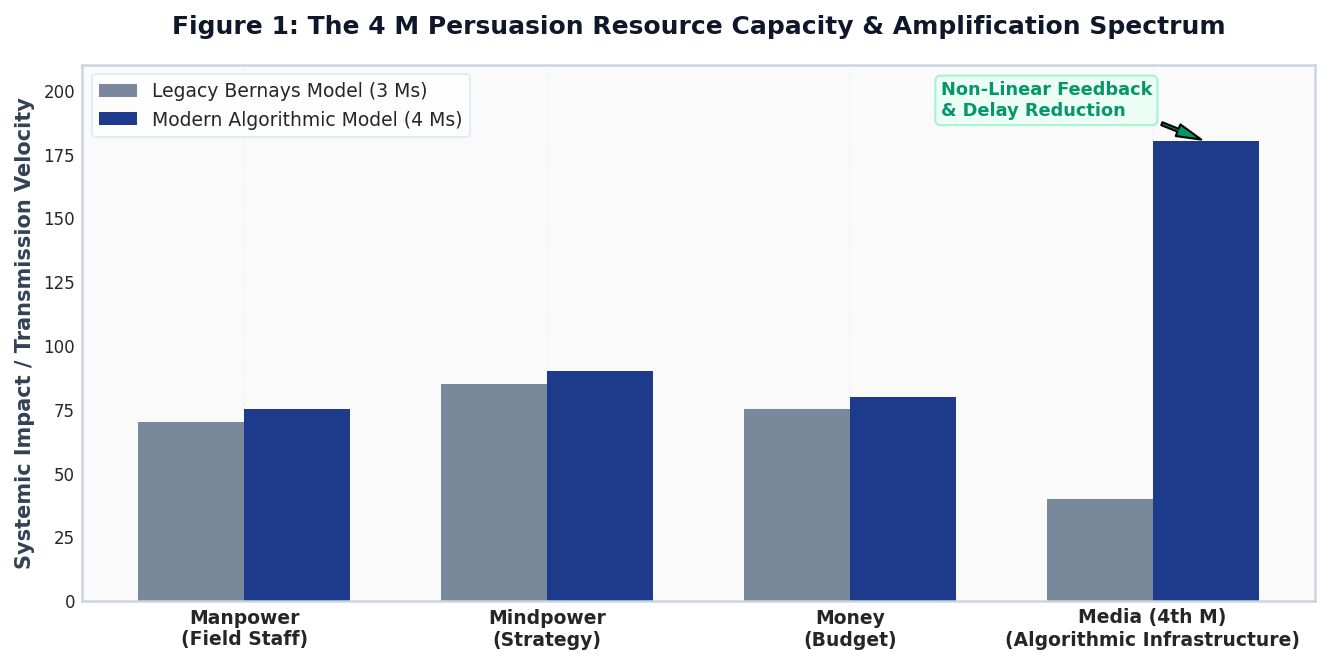}
\caption{Conceptual illustration of the 4M Political Communication Resource Capacity and Amplification Spectrum, highlighting how algorithmic media infrastructure alters transmission dynamics~\cite{bernays1955, ramli2026}.}
\label{fig:f1}
\end{figure}

As illustrated conceptually in Figure~\ref{fig:f1}, adding algorithmic media infrastructure to the traditional resource base may increase transmission speed, network reach, and feedback intensity~\cite{bernays1955, ramli2026}. The magnitude and direction of these effects remain conditional on platform design, audience network structure, message characteristics, and opposition counter-messaging~\cite{woolley2016, ramli2026}.

\section{Structural Mapping: Bernays' Eight Steps vs. S-E-E-D System Dynamics}
To operationalize Dynamic Democratic Consent, Bernays' classic eight-step public relations process must be structurally integrated with the analytical variables of system dynamics~\cite{bernays1955, ramli2026}. Bernays stressed that strategic planning is not a haphazard collection of tricks or gimmicks, but a carefully engineered blueprint requiring top-level leadership integration, realistic goal formulation, and explicit timing calibration~\cite{bernays1955}. This integration establishes a three-phase operational workflow: System Calibration, System Architecture, and Delay and Execution Control~\cite{bernays1955, ramli2026}.

\begin{table}[ht]
\centering
\caption{Structural Mapping: Bernays' Eight Steps vs. S-E-E-D System Dynamics}
\label{tab:bernays_seed_mapping}
\begin{tabular}{p{2.5cm}p{3.2cm}p{3.5cm}p{4.2cm}}
\toprule
\textbf{Phase} & \textbf{Bernaysian PR Step} & \textbf{S-E-E-D Variable} & \textbf{Operational Political Strategy} \\
\midrule
\textbf{I. System Calibration} & 1. Define Objectives~\cite{bernays1955}. & Reference Support Target ($S^*$)~\cite{ramli2026}. & Establishing explicit vote shares and policy targets~\cite{bernays1955, ramli2026}. \\
 & 2. Research Publics~\cite{bernays1955}. & Boundary and Time Lag ($AT_p$)~\cite{ramli2026}. & Auditing baseline sentiment and information delays~\cite{bernays1955, ramli2026}. \\
 & 3. Modify Objectives~\cite{bernays1955}. & Goal vs.\ Capacity Ceiling ($K_{\text{dem}}$)~\cite{ramli2026}. & Adjusting targets to demographic limits~\cite{bernays1955, ramli2026}. \\
\midrule
\textbf{II. System Architecture} & 4. Decide Strategy~\cite{bernays1955}. & Dominance Shift Plan ($R_1 \to B_1$)~\cite{ramli2026}. & Transitioning from voter acquisition to retention~\cite{bernays1955, ramli2026}. \\
 & 5. Set Themes/Symbols~\cite{bernays1955}. & Reinforcing ($R_1$) Triggers~\cite{ramli2026}. & Deploying resonant narratives to drive viral growth~\cite{bernays1955, ramli2026}. \\
 & 6. Blueprint Org.\ \cite{bernays1955}. & 4M Resource Allocation~\cite{ramli2026}. & Structuring staff, budget, and digital media~\cite{bernays1955, ramli2026}. \\
\midrule
\textbf{III. Delay Control} & 7. Chart Timing/Tactics~\cite{bernays1955}. & Perception Delay ($AT_p$) and Elasticity Management~\cite{ramli2026}. & Synchronizing policy delivery with campaign pushes~\cite{bernays1955, ramli2026}. \\
 & 8. Carry Out Tactics~\cite{bernays1955}. & Feedback Loop Execution~\cite{ramli2026}. & Deploying real-time multi-channel communication~\cite{bernays1955, ramli2026}. \\
\bottomrule
\end{tabular}
\end{table}

In Phase I (System Calibration), the political organization defines its reference support target $S^*$ by establishing an explicit intended-support level~\cite{bernays1955, ramli2026}. Step 2 involves auditing internal and external voter segments to map structural boundaries and measure information perception lags ($AT_p$)~\cite{bernays1955, ramli2026}. Step 3 enforces realistic goal modifications, ensuring that target objectives avoid subjective wish fulfillment and remain strictly within the demographic carrying capacity ($K_{\text{dem}}$) of the electorate, thereby preventing political overreach~\cite{bernays1955, ramli2026}.

In Phase II (System Architecture), Step 4 maps the planned dominance shift ($R_1 \to B_1$), structuring campaign phases to transition smoothly from aggressive voter acquisition to coalition stabilization~\cite{bernays1955, ramli2026}. Step 5 designs high-resonance emotional themes and visual symbols that connect with basic human motivations, acting as triggers for self-multiplying reinforcing loops~\cite{bernays1955, entman1993, ramli2026}. Step 6 blueprints the operational organization, aligning the allocation of the 4Ms across internal and external publics to prevent institutional disorganization~\cite{bernays1955, ramli2026}.

In Phase III (Delay and Execution Control), Step 7 establishes precise operational schedules and incorporates Bernays' recommended safety reserve in time, money, and manpower to absorb unforeseen environmental strains, monitor perception lags, and mitigate delay-induced instability~\cite{bernays1955, ramli2026}. Step 8 executes real-time, multi-channel tactical campaigns across algorithmic distribution networks, dynamically adjusting messaging based on continuous sentiment feedback~\cite{bernays1955, ramli2026}.

\section{The S-E-E-D Electoral Dynamics Engine}
The core of Dynamic Democratic Consent rests upon the four fundamental behavior modes defined by the S-E-E-D framework: Snowball dynamics, Equilibrium constraints, Elastic time lags, and Dominance shifts~\cite{ramli2026}.

To ensure analytical clarity across system levels, Table~\ref{tab:variable_taxonomy} provides a selection of mathematical symbols and operational definitions employed within the model.

\begin{table}[htbp]
\centering
\caption{Selected Mathematical Symbols and Operational Definitions}
\label{tab:variable_taxonomy}
\begin{tabular}{p{2.0cm}p{3.5cm}p{4.0cm}p{1.8cm}p{2.2cm}}
\toprule
\textbf{Symbol} & \textbf{Construct Name} & \textbf{Operational Definition} & \textbf{Unit} & \textbf{System Status} \\
\midrule
$S(t)$ & Electorate Support Base & Proportion of target electorate intending to support organization & Percentage & Primary State Stock \\
$P(t)$ & Perceived Support & Information-smoothed perception stock of public alignment & Percentage & Information Stock \\
$T(t)$ & Institutional Trust Stock & Accumulated institutional credibility and civic goodwill & Index ($0$--$100$) & Dynamic State Stock \\
$S^*$ & Reference Support Target & Desired proportion of target electorate intending to support party & Percentage & Target Parameter \\
$K_{\text{dem}}$ & Demographic Capacity & Maximum theoretical ceiling of potential voter base & Percentage & Structural Parameter \\
$K_{\text{effective}}(t)$ & Effective Support Ceiling & Dynamic trust-adjusted attainable support ceiling & Percentage & Dynamic Parameter \\
$\lambda$ & Baseline Accessible Fraction & Minimum accessible fraction of demographic capacity under low trust & Dimensionless & Bounded Parameter ($0 \le \lambda \le 1$) \\
$AT_s$ & Support Adjustment Lag & Average campaign action adjustment time constant & Weeks & Adjustment Parameter \\
$AT_p$ & Perception Delay Constant & Average perception and survey reporting time constant & Weeks & Delay Parameter \\
$g$ & Growth Fractional Rate & Net fractional mobilization accumulation rate & $1/\text{Weeks}$ & Fractional Rate \\
$\alpha$ & Campaign Reaction Gain & Rate at which perceived deviation from target changes support & $1/\text{Weeks}$ & Response Parameter \\
$\beta$ & Support Attrition Rate & Fractional rate of support loss without reinforcement & $1/\text{Weeks}$ & Decay Parameter \\
\bottomrule
\end{tabular}
\end{table}

\begin{figure}[ht]
\centering
\includegraphics[width=0.85\linewidth]{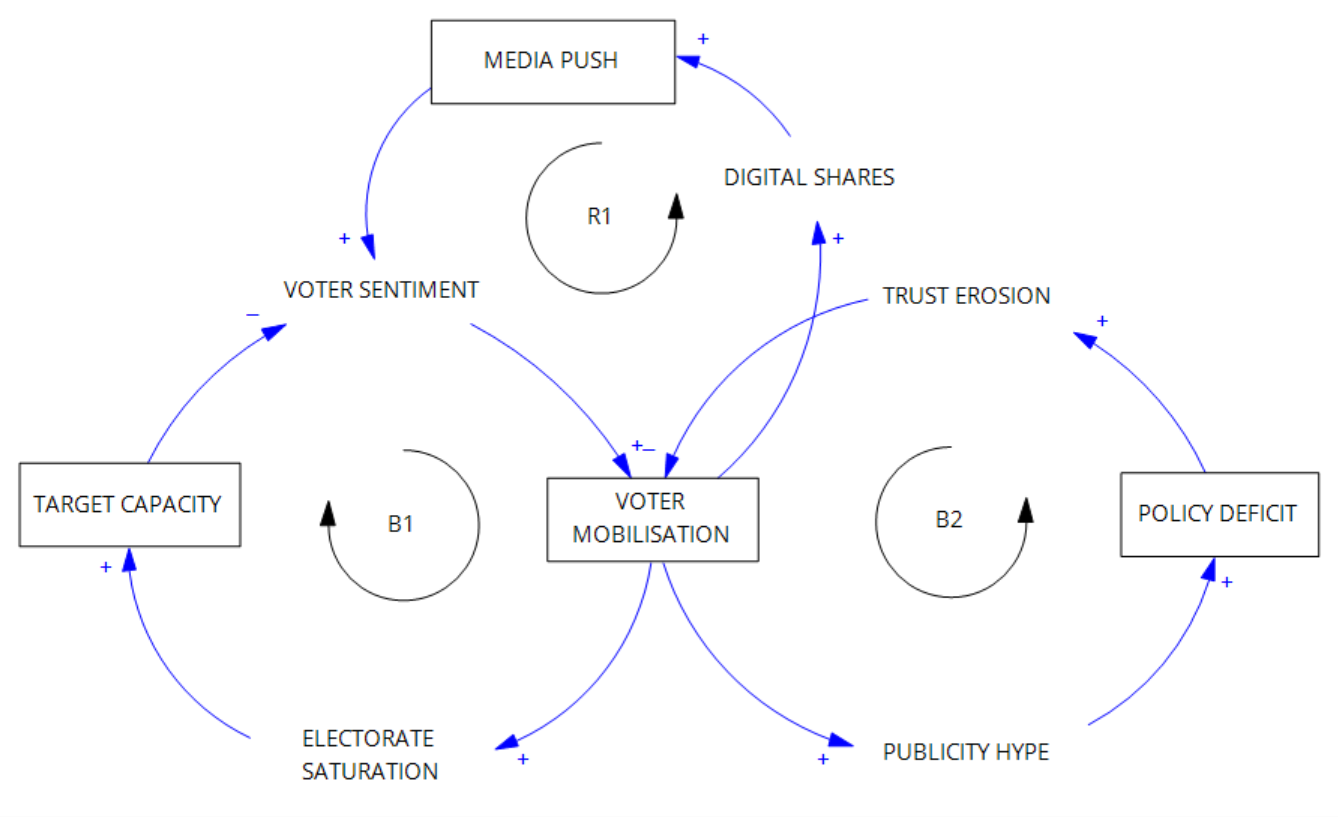}
\caption{System Dynamics Causal Loop Diagram illustrating the interconnected Reinforcing Momentum ($R_1$), Electorate Saturation ($B_1$), and Trust Decay ($B_2$) feedback structures~\cite{bernays1955, ramli2026}.}
\label{fig:cld}
\end{figure}

The causal loop diagram in Figure~\ref{fig:cld} maps three interconnected feedback structures centered on the primary state variable, Electorate Support $S(t)$, defined as the proportion of the target electorate intending to support the political organization~\cite{ramli2026}. The primary reinforcing loop ($R_1$: Viral Momentum) formalizes self-amplifying growth: an intensified \textit{Media Push} elevates \textit{Voter Sentiment}, accelerating the support mobilization rate into $S(t)$, which generates peer-to-peer \textit{Digital Shares} that continuously reinforce media reach~\cite{ramli2026}. Counterbalancing this momentum, the first balancing loop ($B_1$: Saturation Ceiling) incorporates demographic carrying capacity limits ($K_{\text{dem}}$). Here, expanding support $S(t)$ increases \textit{Electorate Saturation}, approaching \textit{Target Capacity} and ultimately suppressing \textit{Voter Sentiment} to stabilize growth along an S-shaped equilibrium trajectory~\cite{ramli2026}. Simultaneously, the second balancing loop ($B_2$: Trust Decay) formalizes Bernays' classic warning regarding unfulfilled public expectations: over-promising without policy delivery destabilizes institutional goodwill~\cite{bernays1955}. In system dynamics terms, scaling $S(t)$ generates unearned \textit{Publicity Hype} that widens the \textit{Policy Deficit} when unbacked by substantive governance deliverables. This expanding deficit induces \textit{Trust Erosion}, directly dampening long-term support $S(t)$ and increasing vulnerability to dynamic overshoot and rapid support decline~\cite{bernays1955, hetherington2005, pierson1993, ramli2026}.

The following equations represent modular reference-mode formulations rather than a single fully integrated predictive model. Each formulation isolates a particular dynamic mechanism, including reinforcing growth, goal-seeking adjustment, delayed corrective response, trust-adjusted saturation, or institutional-trust accumulation. A subsequent stock-and-flow implementation would combine these mechanisms and evaluate their joint behavior under explicit parameter assumptions.

Snowball dynamics ($S$) are powered by positive, self-reinforcing loops ($R_1$) that drive exponential mobilization, digital fundraising compounding, and viral narrative adoption~\cite{sterman2000, ramli2026}. The absolute rate of change in political support $S(t)$ over time $t$ is expressed as:
\begin{equation}
\frac{dS}{dt} = g S(t)
\end{equation}
where $g$ represents the constant fractional growth rate~\cite{sterman2000, ramli2026}. Unconstrained $R_1$ loops produce sharp exponential curves where small initial gains rapidly compound into political momentum~\cite{ramli2026}.

\begin{figure}[ht]
\centering
\includegraphics[width=0.85\linewidth]{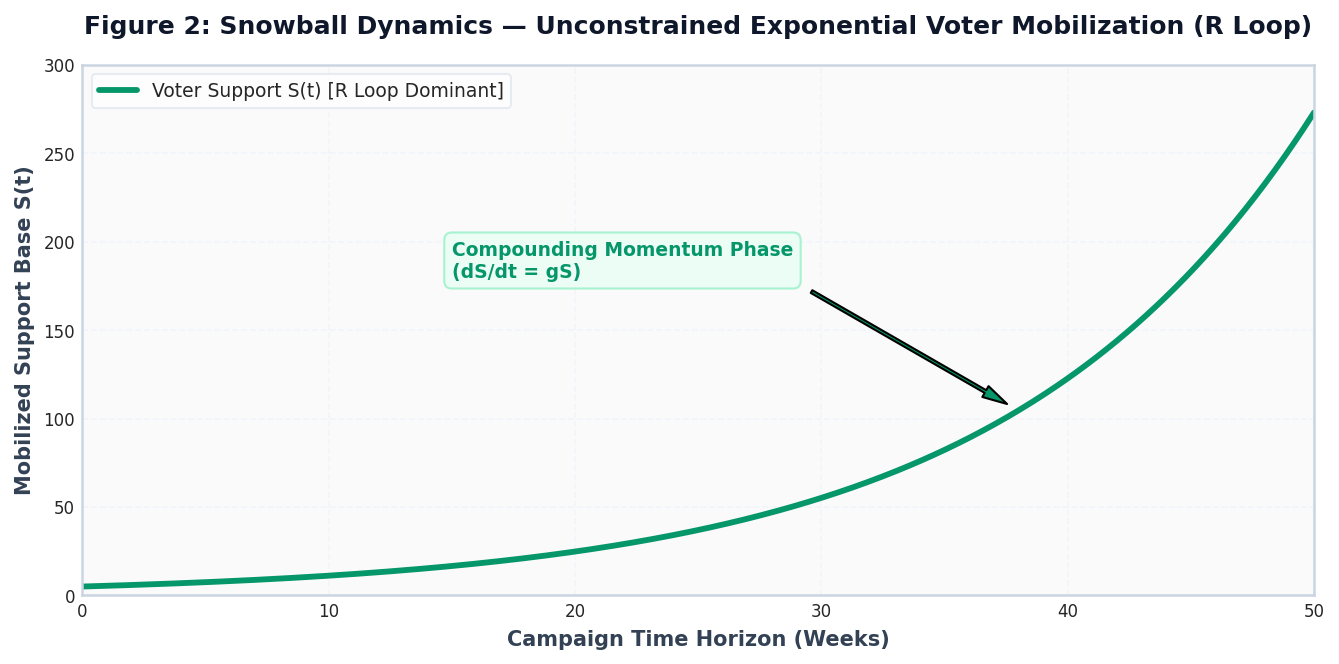}
\caption{Simulated behavior mode of unconstrained exponential voter mobilization driven by a dominant positive reinforcing feedback loop ($R_1$)~\cite{ramli2026}.}
\label{fig:f2}
\end{figure}

As depicted in Figure~\ref{fig:f2}, unconstrained snowball dynamics generate rapid growth trajectory curves that begin gradually before accelerating over a fixed campaign horizon~\cite{ramli2026}. Political strategists leverage this mode during early campaign phases to build grassroots excitement, donor momentum, and brand prominence~\cite{bernays1955, ramli2026}.

However, political growth never occurs in an isolated vacuum; it is bounded by Equilibrium constraints ($E$) enforced by negative balancing loops ($B_1$)~\cite{sterman2000, ramli2026}. In the absence of an explicit perception delay, goal-seeking support adjustment toward a target baseline ($S^*$) is governed by a first-order adjustment equation:
\begin{equation}
\frac{dS}{dt} = \frac{S^* - S(t)}{AT_s}
\end{equation}
where $S^*$ is the desired target goal and $AT_s$ represents the support adjustment time constant~\cite{sterman2000, ramli2026}. Equation 2 produces smooth, monotonic asymptotic convergence toward $S^*$.

Elastic behavior ($E$) and underdamped oscillations emerge when decision-makers respond to a delayed or smoothed perception of support, $P(t)$, rather than its true instantaneous value~\cite{sterman2000}. Mathematically, perceived support accumulates as an information stock according to:
\begin{equation}
\frac{dP}{dt} = \frac{S(t) - P(t)}{AT_p}
\end{equation}
where $AT_p$ represents the perception delay time constant~\cite{sterman2000}. Campaign action and support adjustments then react to the perceived gap $(S^* - P(t))$, establishing a second-order feedback response:
\begin{equation}
\frac{dS}{dt} = \alpha \left( S^* - P(t) \right) - \beta S(t)
\end{equation}
where $\alpha$ is the campaign reaction gain and $\beta$ represents natural support attrition. At steady-state equilibrium ($\frac{dP}{dt} = 0, \frac{dS}{dt} = 0$), $P_{\text{eq}} = S_{\text{eq}}$, yielding:
\begin{equation}
S_{\text{eq}} = \frac{\alpha}{\alpha + \beta} S^*
\end{equation}
For positive reaction gain $\alpha$ and natural attrition $\beta$, the steady-state support level settles at $S_{\text{eq}} = \frac{\alpha}{\alpha + \beta} S^*$. Thus, natural support attrition produces a persistent steady-state shortfall relative to the reference target unless corrective response is sufficiently strong ($\alpha \gg \beta$) or an additional baseline support inflow is present.

\begin{figure}[ht]
\centering
\includegraphics[width=0.85\linewidth]{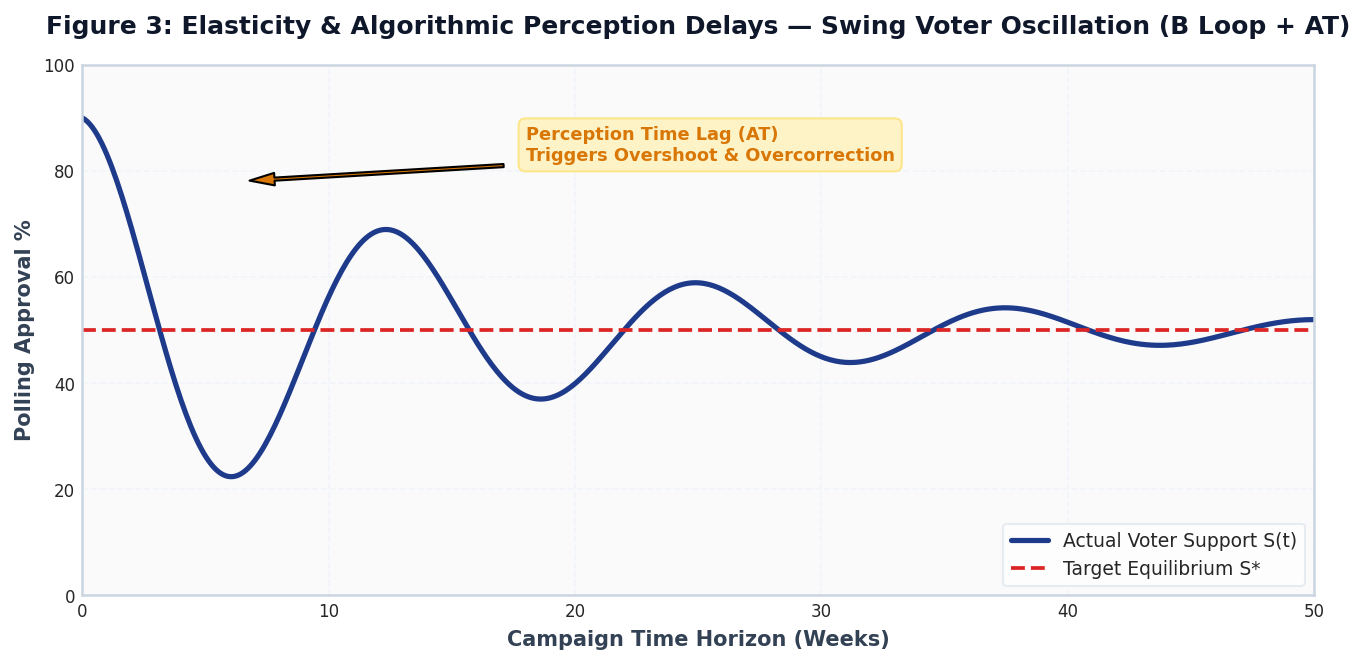}
\caption{Simulated behavior mode showing polling oscillations generated by delayed balancing feedback mechanisms ($B_1 + AT_p$)~\cite{ramli2026}.}
\label{fig:f3}
\end{figure}

Damped oscillatory behavior occurs when the reaction gain, attrition rate, and perception delay satisfy:
\begin{equation}
\left( \beta AT_p - 1 \right)^2 < 4 \alpha AT_p
\end{equation}
This mathematical condition indicates that delayed corrective response must be sufficiently strong relative to the system's effective damping. Figure~\ref{fig:f3} illustrates underdamped adjustment generated formally by delayed perception and corrective response. In applied settings, aggressive responses to delayed polling information may contribute to overshooting and may also interact with mechanisms such as voter fatigue and opposition counter-mobilization~\cite{bernays1955, ramli2026}.

Figure~\ref{fig:f3} illustrates how unmanaged perception time constants ($AT_p$) induce underdamped oscillatory behavior~\cite{ramli2026}. When campaign strategists react to delayed polling data by escalating messaging expenditures, they inadvertently induce overshooting, triggering swing-voter fatigue and counter-mobilization from opposition coalitions~\cite{bernays1955, ramli2026}.

Real-world political competition is non-linear, characterized by coupled feedback loops competing for dominance over the same demographic voter base~\cite{sterman2000, ramli2026}. As a political party expands its reach, control naturally shifts from the accelerating reinforcing loop to the constraining balancing loop, generating an S-shaped growth trajectory governed by:
\begin{equation}
\frac{dS}{dt} = g S(t) \left( 1 - \frac{S(t)}{K_{\text{effective}}(t)} \right)
\end{equation}
where $K_{\text{effective}}(t)$ represents the dynamic effective support ceiling, defined as:
\begin{equation}
K_{\text{effective}}(t) = K_{\text{dem}} \cdot f(T(t)) = K_{\text{dem}} \left[ \lambda + (1 - \lambda)\frac{T(t)}{100} \right], \quad 0 \le \lambda \le 1
\end{equation}
where $K_{\text{dem}}$ is the structural demographic ceiling, $T(t)$ is the dynamic institutional trust stock normalized to the bounded interval $0 \le T(t) \le 100$, and $\lambda$ represents the baseline fraction of demographic capacity accessible even under low trust conditions~\cite{hetherington2005, sterman2000, ramli2026}. Lower values of $\lambda$ imply greater sensitivity of effective support capacity to institutional trust, whereas $\lambda = 1$ represents a limiting case in which trust does not alter demographic support capacity. The function $f(T(t))$ is bounded between zero and one, positive, and monotonically increasing in institutional trust.

\begin{figure}[ht]
\centering
\includegraphics[width=0.85\linewidth]{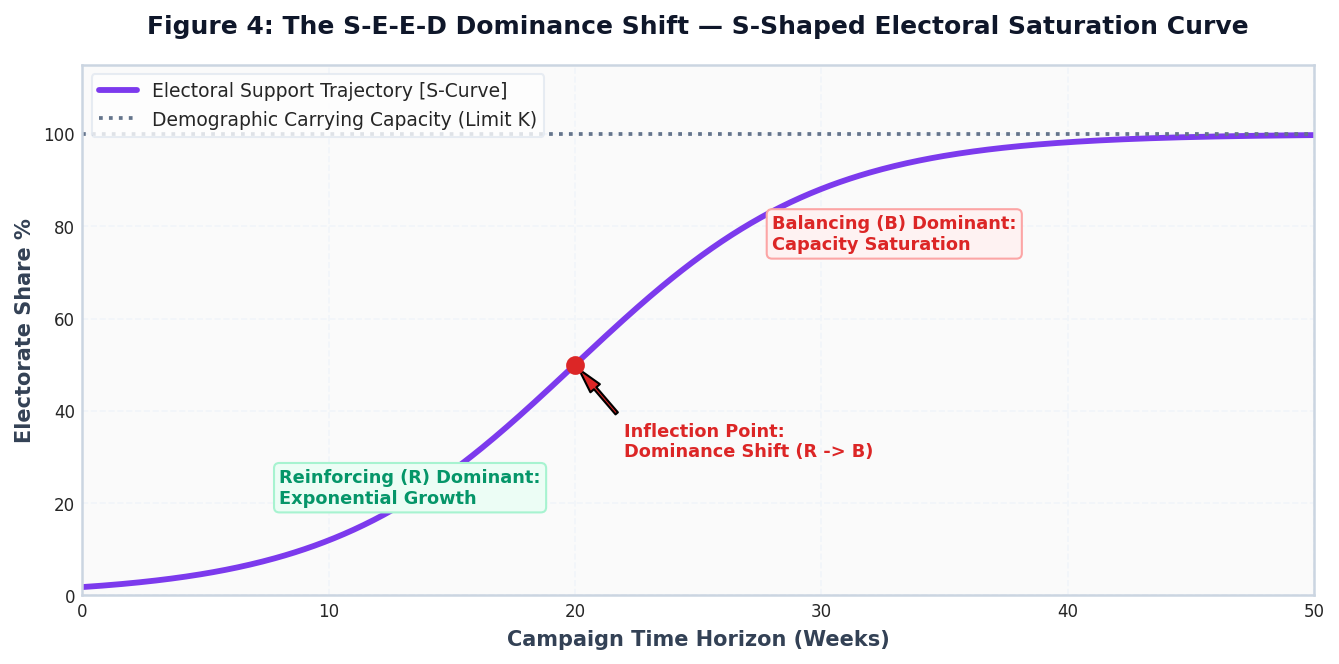}
\caption{Simulated behavior mode of an S-shaped electoral saturation trajectory illustrating the shift in loop dominance from $R_1$ to $B_1$ at the inflection point~\cite{ramli2026}.}
\label{fig:f4}
\end{figure}

Figure~\ref{fig:f4} isolates the structural inflection point where the dominant driver of political performance transfers from the accelerating snowball loop to the stabilizing balancing constraint~\cite{sterman2000, ramli2026}. In system dynamics terminology, a dominance shift represents a transfer of structural influence between coupled feedback channels rather than a physical transformation of one loop into another~\cite{sterman2000}. Prior to this inflection point, resource abundance enables exponential voter acquisition; beyond it, market saturation sets in, requiring strategists to shift from acquisition tactics to retention policies~\cite{bernays1955, ramli2026}.

\section{The 4-Step Political Diagnostic Pipeline}
To operationalize dynamic insights during active political campaigns, analysts utilize the \textbf{Four-Step System Diagnostic Blueprint}~\cite{ramli2026}. This pipeline enables campaign strategists to systematically audit active performance, isolate underlying loop structures, design high-leverage structural interventions, and project modified dynamic trajectories~\cite{bernays1955, sterman2000, ramli2026}.

The diagnostic pipeline operates sequentially through four interconnected stages:
\begin{equation}
\text{[1. Current Trajectory]} \longrightarrow \text{[2. Loop Mapping]} \longrightarrow \text{[3. Policy Design]} \longrightarrow \text{[4. Projected Shift]}
\end{equation}

In Step 1 (Current Trajectory Audit), analysts plot active voter approval, digital engagement, and fundraising data over time to determine whether the political baseline is spiking, flatlining, oscillating, or declining~\cite{ramli2026}. In Step 2 (Loop Mapping), analysts evaluate the structural feedback mechanisms underlying the observed trajectory, identifying whether movements are driven by viral $R_1$ loops, institutional $B_1$ resistance, or unmanaged time delays ($AT_p$)~\cite{ramli2026}. In Step 3 (Policy Design), strategists formulate targeted interventions guided by Bernays' core principle of ``deeds before words'' (reconstructing core policy deliverables rather than increasing superficial publicity)~\cite{bernays1955, ramli2026}. In Step 4 (Projected Shift), analysts examine the simulated interval in which the intervention may alter relative loop dominance and support stabilization~\cite{ramli2026}.

\begin{figure}[ht]
\centering
\includegraphics[width=0.85\linewidth]{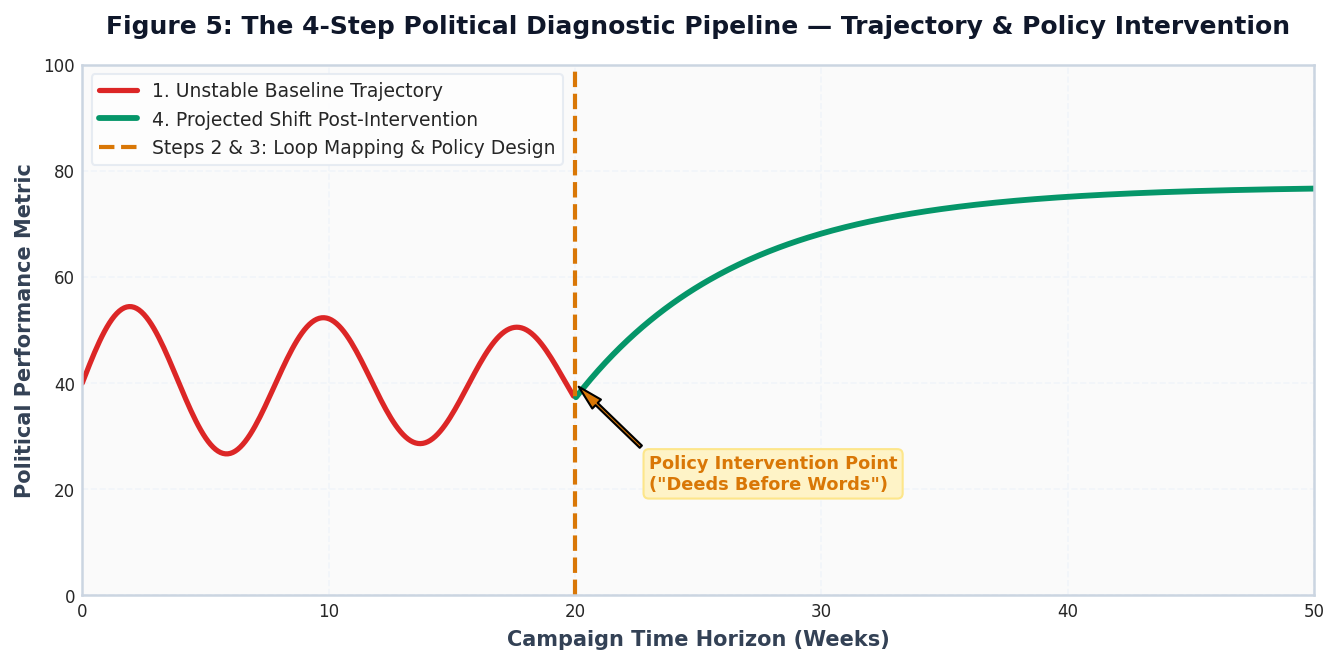}
\caption{Conceptual illustration of performance trajectory transformation following a high-leverage structural policy intervention at $t=20$~\cite{bernays1955, ramli2026}.}
\label{fig:f5}
\end{figure}

Figure~\ref{fig:f5} demonstrates the practical impact of the diagnostic pipeline~\cite{ramli2026}. In the conceptual intervention scenario, the policy change at $t=20$ is represented by a reduction in perception delay $AT_p$ and a moderation of excessive campaign reaction gain $\alpha$, thereby dampening oscillatory overcorrection. Simultaneously, improvements in substantive policy performance lower support attrition $\beta$ and elevate the trust-formation rate, expanding $K_{\text{effective}}(t)$. Prior to intervention ($t < 20$), the political trajectory displays unstable polling oscillations driven by unmanaged perception delays~\cite{ramli2026}. Under these illustrative assumptions, these combined mechanisms allow the trajectory to transition from unstable polling oscillations to a smooth, goal-seeking curve that converges toward a higher sustainable support level~\cite{bernays1955, ramli2026}.

\section{Systemic Failure Modes and Illustrative Reference Mode Matching}
When political parties rely on open-loop communication while ignoring structural feedback dynamics, campaign execution frequently degrades into two primary failure pathologies: the Unbacked Performance Gap State and the Policy Resistance Plateau~\cite{bernays1955, pierson1993, ramli2026}.

The Unbacked Performance Gap failure occurs when a political organization uses aggressive public relations campaigns to mask unfulfilled policy promises or administrative backlogs~\cite{bernays1955, hetherington2005, ramli2026}. Institutional trust accumulates as a dynamic state stock $T(t)$ bounded between 0 and 100, driven by the net rate of trust formation relative to trust erosion:
\begin{equation}
\frac{dT}{dt} = \text{Trust Formation} - \text{Trust Erosion}
\end{equation}
In simulation execution, trust formation and erosion flows are subject to boundary constraints maintaining $0 \le T(t) \le 100$. Trust formation increases with observable policy delivery, transparent communication, and expectation alignment, whereas trust erosion increases with the accumulated gap between communicated claims and observable policy performance~\cite{bernays1955, hetherington2005}. In system dynamics terms, publicity creates a short-term surge in political support that temporarily overshoots actual performance~\cite{ramli2026}. However, this unearned publicity tends to progressively erode $T(t)$, contracting the effective support ceiling $K_{\text{effective}}(t)$ and creating a fragile resource base~\cite{bernays1955, hetherington2005, ramli2026}. As declining institutional trust contracts the effective support ceiling below existing support levels, the system experiences accelerated political-support decay toward a lower equilibrium~\cite{ramli2026}.

\begin{figure}[ht]
\centering
\includegraphics[width=0.85\linewidth]{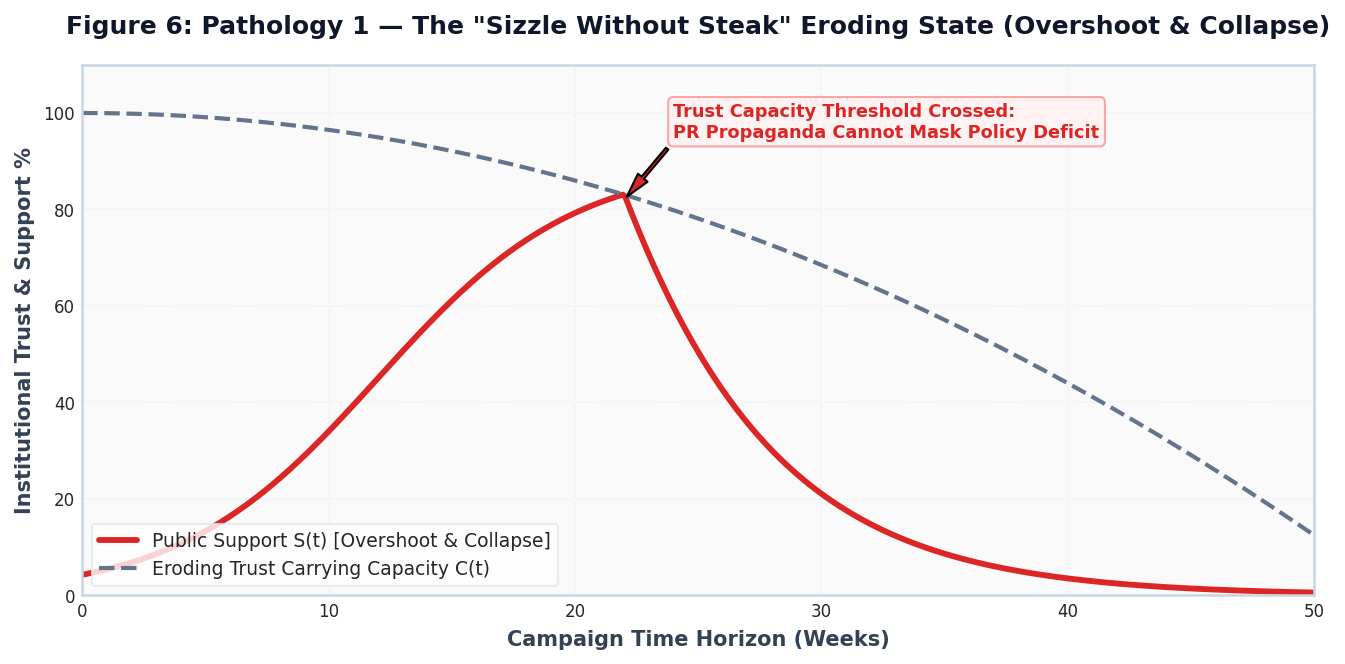}
\caption{Simulated behavior mode of the Unbacked Performance Gap State, illustrating how excessive PR messaging contracts effective support capacity and accelerates support decay~\cite{bernays1955, ramli2026}.}
\label{fig:f6}
\end{figure}

Figure~\ref{fig:f6} illustrates accelerated political-support decay produced when declining institutional trust contracts the effective support ceiling below the existing support level~\cite{ramli2026}. While short-term publicity generates a temporary rise in approval, the underlying dynamic stock of institutional trust decays continuously~\cite{bernays1955, hetherington2005, ramli2026}. Once effective support capacity contracts below active support levels, the system experiences a drop in approval that communication adjustments alone cannot arrest~\cite{bernays1955, ramli2026}.

The second failure mode, the Policy Resistance Plateau (Stagnant State), occurs when a campaign escalates media advertising expenditure without resolving underlying voter grievances~\cite{pierson1993, ramli2026}. The campaign's aggressive push activates unrecognized, compensating balancing loops ($B_1, B_2$), such as voter fatigue, media skepticism, and opposition counter-mobilization, that actively push back against campaign outreach~\cite{sterman2000, ramli2026}.

\begin{figure}[ht]
\centering
\includegraphics[width=0.85\linewidth]{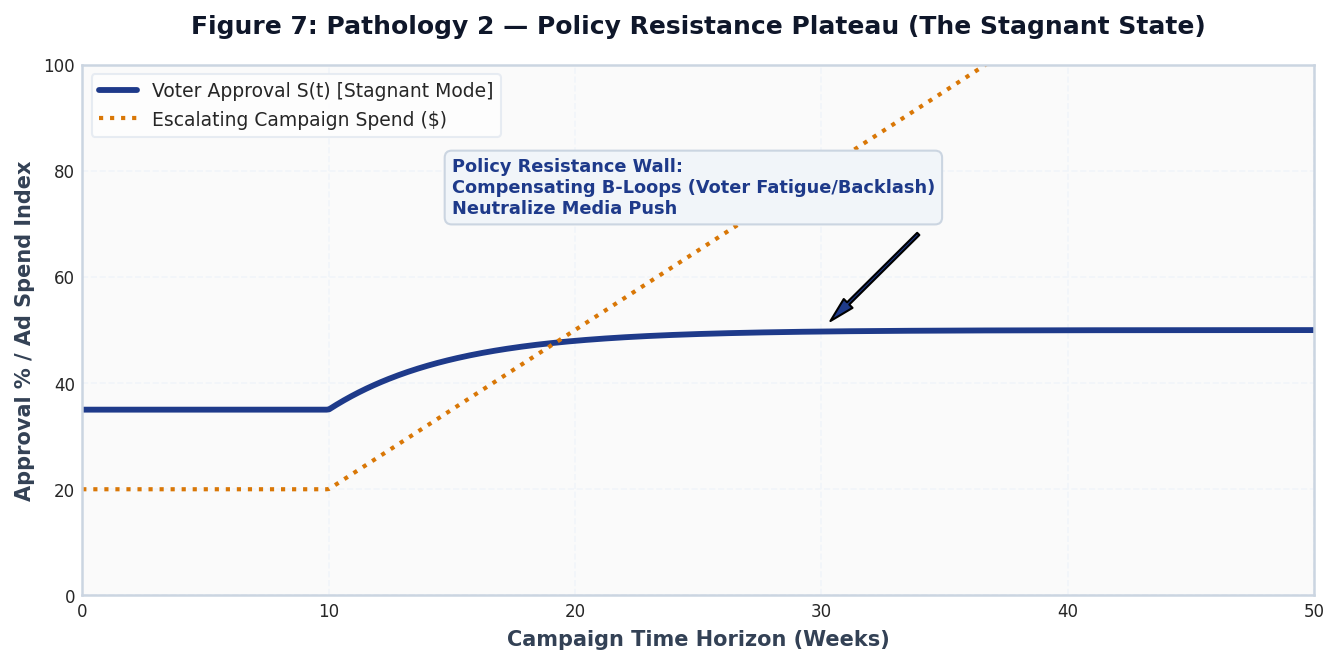}
\caption{Simulated behavior mode of the Policy Resistance Plateau, showing how escalating campaign advertising expenditure encounters compensating balancing loops~\cite{ramli2026}.}
\label{fig:f7}
\end{figure}

As shown in Figure~\ref{fig:f7}, increasing advertising spend ($Money$) against compensating balancing loops yields diminishing returns, locking political approval into an underachieving plateau~\cite{ramli2026}. To break through this barrier, strategists must reduce the strength of compensating balancing mechanisms by resolving the root causes of voter dissatisfaction before scaling external communications~\cite{bernays1955, pierson1993, ramli2026}.

To evaluate whether computer simulation models of political behavior capture essential real-world dynamics, system dynamicists execute Reference Mode Verification~\cite{forrester1961, sterman2000, ramli2026}. Rather than claiming statistical estimation or econometric curve fitting, reference mode matching assesses whether a model's internal loop structures qualitatively replicate fundamental behavioral modes, phase lags, inflections, and rhythms observed in historical contexts~\cite{sterman2000, ramli2026}.

\begin{figure}[ht]
\centering
\includegraphics[width=0.85\linewidth]{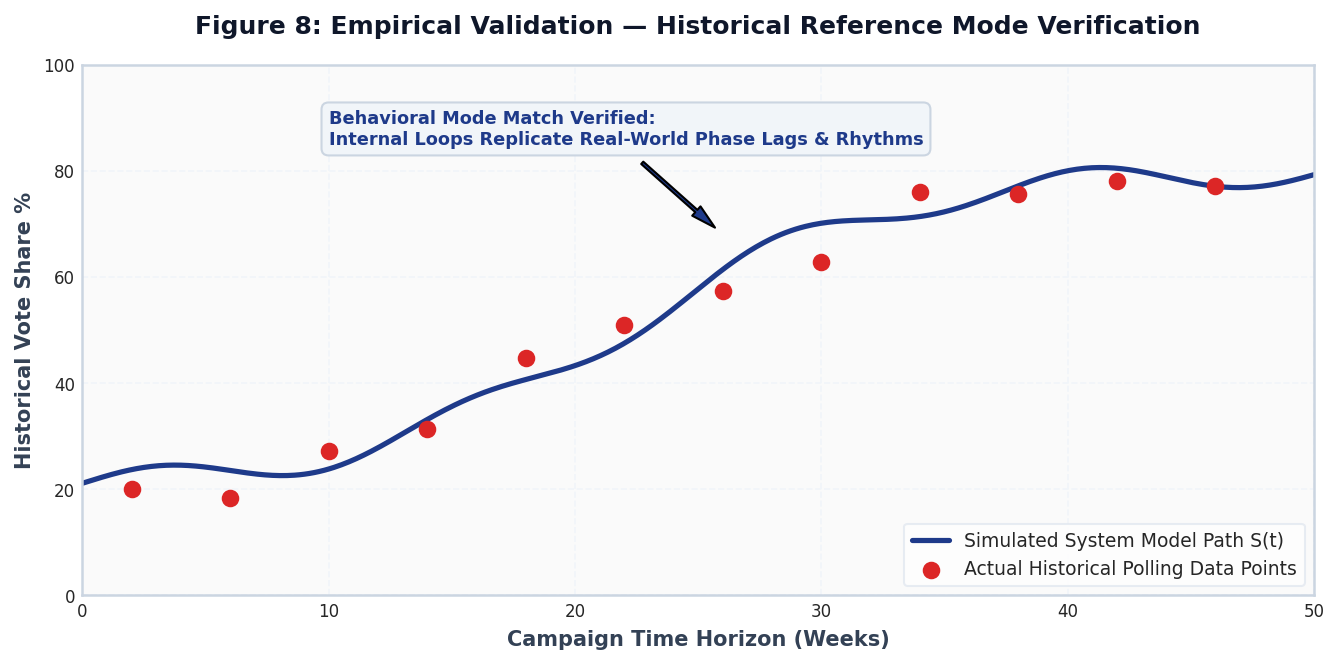}
\caption{Synthetic benchmark illustrating qualitative reference-mode comparison between a simulated trajectory and a hypothetical observational pattern~\cite{ramli2026}.}
\label{fig:f8}
\end{figure}

Figure~\ref{fig:f8} illustrates how a simulated trajectory may be compared with a reference pattern in terms of broad behavioral characteristics, including phase lag, growth rates, inflection points, and equilibrium plateaus~\cite{ramli2026}. Because the benchmark is synthetic, the comparison demonstrates the model-evaluation procedure rather than empirical correspondence with a particular election. Full parameter estimation, formal statistical calibration, and out-of-sample empirical testing against specific historical election datasets remain subjects for subsequent empirical research.

\section{Ethical Safeguards, Strategic Alignment, and Institutional Resilience}
The deployment of Dynamic Democratic Consent introduces significant ethical considerations regarding democratic integrity and institutional stability~\cite{bernays1955, norris2011, ramli2026}. Political organizations routinely leverage strategic framing architectures to reframe administrative adjustments in constructive operational terms~\cite{bernays1955, entman1993}. Within DDC, principles such as \textit{Proactive Consensus Alignment} (systematic public outreach calibration), \textit{Strategic Perception Re-indexing} (contextualizing policy implementation timelines), and \textit{Expectation Management} (modifying public performance baselines) serve as structured mechanisms to harmonize institutional outputs with civic sentiment~\cite{bernays1955, ramli2026}.

In this framework, expectation alignment refers to transparent communication regarding implementation constraints, timelines, uncertainties, and measurable outcomes. It explicitly excludes withholding material information, fabricating achievements, suppressing criticism, or presenting delayed policy delivery as completed performance~\cite{bernays1955, norris2011}. To ensure democratic integrity, DDC mechanisms must operate within four normative safeguards:
\begin{enumerate}
    \item \textbf{Transparency}: Mandatory disclosure of message provenance, funding sources, campaign sponsorship, and algorithmic targeting parameters~\cite{norris2011, woolley2016}.
    \item \textbf{Public Autonomy}: Protection of citizen deliberative autonomy by prohibiting deceptive, covert, or psychologically coercive micro-targeting~\cite{bennett2008, zaller1992}.
    \item \textbf{Data Ethics}: Strict prohibitions against predatory behavioral profiling, unauthorized data exploitation, or discriminatory voter suppression tactics~\cite{woolley2016}.
    \item \textbf{Contestability}: Institutional preservation of public space for open criticism, independent fact-checking, and vigorous political opposition~\cite{norris2011, pierson1993}.
\end{enumerate}

Informational distortion may generate short-term communication advantages, but the DDC framework proposes that sustained divergence between claims and observable policy outcomes increases long-run trust erosion and systemic fragility~\cite{bernays1955, hetherington2005, ramli2026}. As established in the performance gap pathology, substituting communication strategies for core policy execution progressively decays the system's underlying trust stock $T(t)$~\cite{bernays1955, hetherington2005, pierson1993}. When external environmental shocks or institutional disruptions occur (such as economic volatility or geopolitical crises), information suppression amplifies systemic delays, triggering severe overcorrection and public disengagement once objective outcomes manifest~\cite{norris2011, ramli2026}.

Sustained political stability requires institutional transparency and policy alignment~\cite{bernays1955, hetherington2005, norris2011}. When political organizations establish open communication channels, address operational bottlenecks promptly, and prioritize substantive deliverables over messaging volume, they protect and strengthen the institutional trust stock $T(t)$~\cite{bernays1955, pierson1993, ramli2026}. Ultimately, the DDC framework proposes that resilient democratic governance depends partly on maintaining alignment between institutional policy delivery, transparent communication, and public expectations~\cite{bernays1955, norris2011, ramli2026}.

\subsection{Theoretical and Methodological Limitations}
While the DDC framework provides a structured system-dynamics mapping of political communication, several boundary limitations must be acknowledged:
\begin{enumerate}
    \item \textbf{Conceptual and Non-Empirical Status}: The mathematical formulations and simulation figures presented serve an exploratory function to clarify feedback behavior modes. The model parameters have not been econometrically estimated or calibrated against a specific historical election dataset.
    \item \textbf{Voter Homogeneity Assumption}: The model aggregates public opinion into state stocks ($S(t), P(t)$), omitting micro-level heterogeneity across demographic, ideological, or regional voter sub-populations.
    \item \textbf{Simplified Institutional Structure}: The framework focuses primarily on the single-organization feedback loop, simplifying multi-party strategic interactions, parliamentary legislative bargaining, and electoral institutional rules.
    \item \textbf{Contextual Media Variability}: The model treats algorithmic media infrastructure as an abstract dynamic multiplier, simplifying differences across platform architectures, national regulatory environments, and content moderation regimes.
    \item \textbf{Endogeneity and Causal Identification}: Communication strategy, policy performance, public support, and institutional trust influence one another simultaneously. The present framework does not identify causal effects or resolve endogeneity among these variables.
    \item \textbf{Functional-Form Dependence}: The simulated behavior depends on selected functional forms, including logistic support growth and a linear trust-capacity relationship. Alternative nonlinear specifications may produce different trajectories and dominance conditions.
    \item \textbf{Behavioral and Measurement Validity}: Electorate support, perceived support, and institutional trust are treated as measurable aggregate constructs, but their empirical indicators may vary across polling instruments, electoral contexts, and time. Subsequent research should evaluate construct validity and measurement invariance before estimating model parameters.
\end{enumerate}

\section{Conclusion}
This paper has proposed the Dynamic Democratic Consent (DDC) framework, integrating selected principles from Bernays' account of public relations with the S-E-E-D behavior modes of system dynamics. The framework organizes political support dynamics around reinforcing momentum, saturation constraints, delayed adjustment, changing loop dominance, policy feedback, and institutional trust. It also extends the conventional resource categories of manpower, mindpower, and money by treating algorithmic media infrastructure as a conditional fourth resource whose effects depend on platform architecture, audience networks, contextual noise, and counter-messaging.

The framework's central implication is that communication intensity alone cannot reliably sustain political support. When communicated expectations diverge from policy delivery, short-term mobilization may coexist with longer-term trust erosion and declining effective support capacity. Conversely, interventions addressing policy performance, transparency, and feedback delays may produce more durable outcomes than additional message amplification.

The present study is conceptual and does not claim predictive or causal validation. Its figures represent simulated or synthetic behavior modes intended to clarify possible dynamic mechanisms. Future research should specify a complete stock-and-flow model, estimate parameters using named electoral datasets, test sensitivity to alternative assumptions, compare competing model structures, and evaluate the framework across political and institutional contexts. Such work would determine whether the proposed feedback architecture provides explanatory or predictive value beyond existing theories of political communication, public opinion, and policy feedback.

\end{document}